\documentclass{an}

\usepackage{graphicx}

\begin{document}

\Pagespan{0}{}
\Yearpublication{Year of publication}
\Yearsubmission{Year of submission}
\Month{Month of publication}
\Volume{Volume}
\Issue{Issue}
\DOI{DOI}
\title{Low mass visual binaries in the solar neighbourhood:\\
 The case of HD\,141272\thanks{Based on observations
obtained on La Silla in ESO programs 77.C-0572(A) and Calar Alto project number F06-3.5-016.}}
\author{
    T. Eisenbeiss\inst{1} \thanks{E-mail: eisen@astro.uni-jena.de}
    \and A. Seifahrt\inst{1,2}
    \and M. Mugrauer\inst{1}
    \and T.~O.~B. Schmidt\inst{1}
    \and R. Neuh\"auser\inst{1}
    \and T. Roell\inst{1}}

\institute{Astrophysikalisches Institut und Universit\"ats-Sternwarte Jena, Schillerg\"asschen 2-3, 07745 Jena, Germany \and European Southern Observatory, Karl-Schwarzschild-Str. 2, 85748, Garching, Germany}

\received{15.08.06}

\accepted{29.03.07}

\publonline{\ldots}


\abstract{We search for stellar and substellar companions of young nearby stars to investigate
stellar multiplicity and formation of stellar and substellar companions. We detect common
proper-motion companions of stars via multi-epoch imaging. Their companionship is finally confirmed
with photometry and spectroscopy. Here we report the discovery of a new  co-moving ($13\,\sigma$)
stellar companion $\sim$ 17.8\,arcsec (350\,AU in projected separation) north of the nearby star
HD\,141272 (21\,pc). With EMMI/NTT optical spectroscopy we determined the spectral type of the
companion to be M3$\pm$0.5V. The derived spectral type as well as the near infrared photometry of
the companion are both fully consistent with a $0.26^{+0.07}_{-0.06} M_{\odot}$ dwarf located at
the distance of HD\,141272 (21\,pc). Furthermore the photometry data rules out the pre-main
sequence status, since the system is consistent with the ZAMS of the Pleiades.}

\maketitle

\section{Introduction}
HD\,141272 is a nearby G8 dwarf with a mass of \linebreak $0.83_{-0.03}^{+0.07}$\,$M_{\odot}$ (Nordstr\"om et al.
2004) located in the constellation Serpens Caput ($\alpha_{J2000.0} = 15^h\,48^m\,09.4^s$,
$\delta_{J2000.0}= +01^{\circ}\,34'\,18''$). Its proper motion ($\mu_{\alpha}\cos\delta=-176.19 \pm
1.08$\,mas/yr, $\mu_{\delta}=-166.72 \pm 1.13\,$mas/yr) and parallax ($\pi = 46.84 \pm 1.05$\,mas,
i.e. 21\,pc) are both well determined by the European astrometry satellite \textsl{Hipparcos}
(Perryman et al. 1997). While Montes et al. (2001) list HD\,141272 as a member of the \textsl{Local
association} with an age of $\sim 120\,$Myr (Mart{\'{\i}}n et al. 2001), Fuhrmann (2004) suggested that
this star belongs to the young \textsl{Her-Lyr moving group}, according to its UV-velocities.  The
age of some Her-Lyr members is estimated by Fuhrmann (2004) to approximately 100\,Myr (e.g. HR\,857,
HD\,82443, HD\,113449 and HR\,5829) which recently reached their main sequence position, while others
seemed to be older than $\sim 200$\,Myr \linebreak (Fuhrmann 2004). Also Fuhrmann (2004) argued that
HD\,141272, with an effective temperature of $T_{eff} = (5270 \pm 80)\,$K, an absolute bolometric
magnitude $M_{bol}= (5.54 \pm 0.07)\,$mag and metallicity of $[Fe/H]= (-0.08 \pm 0.07)\,$dex
appears slightly too bright for its main sequence position, indicating that it might be non single
or young.

On the other hand Gaidos, Henry \& Henry (2000) measured a Fe corrected Li-equivalent width of
$W_{6708}=3.9 \pm 1.9\,$m\AA$\ $and a rotational velocity of \linebreak $v \sin i \approx 4.0\,$km/s, which
might be too small for a \linebreak 100\,Myr old star. Furthermore Chen et al. (2005) observed HD\,141272 using
the infrared space telescope \textsl{Spitzer} and did not find any IR-excess at 24$\,\mu$m and
70$\,\mu$m indicating that HD\,141272 is not surrounded by an optically thick disk.

Finally L\'opez-Santiago et al. (2006) revised the list of Her-Lyr members and candidates of
Fuhrmann (2004) and classified HD\,141272 as an doubtful member, due to its lithium depletion.

In our program we search for companions to Her-Lyr members and candidates and first results are
presented here. We found a co-moving companion of \linebreak HD\,141272 by a combination of archival first
epoch images and recent observations. We present our imaging, the  astrometric data and reduction
techniques in section \ref{data} and \ref{data2}, followed by a description of the spectroscopic
and photometric analysis of the new companion in section \ref{photo}. The results are discussed in
section \ref{concl}.

\section{Archival first epoch data}\label{data}
\begin{figure}
\resizebox{\hsize}{!}{\includegraphics{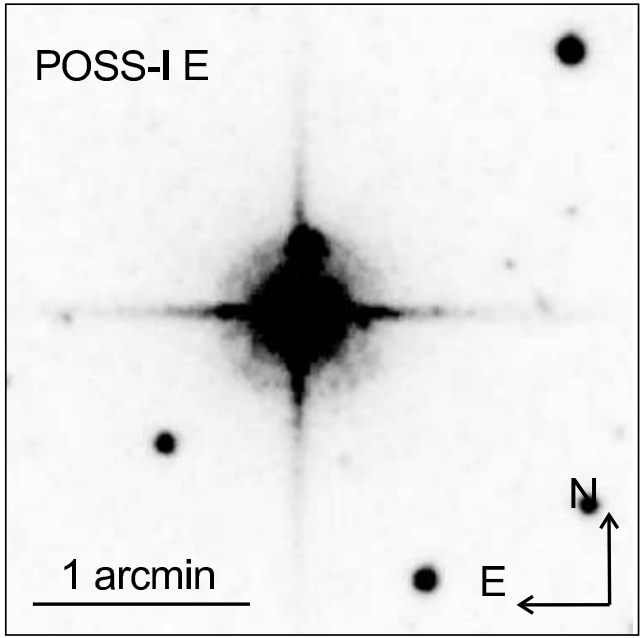}} \caption{POSS-I E image of HD\,141272 from 17 June
1950. The star is located at $\alpha_{J2000.0}= 15^h\,48^m\,09.4^s$, $\delta_{J2000.0}=
+01^{\circ}\,34'\,18''$. A faint object is located in the north of HD\,141272, which is hardly
recognizable due to the diffraction spikes of the primary star induced by saturation. With a pixel
size of $\sim 10\,$microns the pixel scale of the plate is $\sim 6.72\,$arcsec/pixel.}
\label{bild2}
\end{figure}

Astrometry is an effective method to find companions of stars, by comparing two images taken with
sufficiently long epoch difference. In order to find late-type stellar and substellar objects, we
concentrate our search on companions of young stars. Young objects are still in contraction and are
brighter than older  objects of the same mass hence, low mass objects are easier to detect.

We found HD\,141272 in three epochs of the Su\-per\-COS\-MOS-Sky-Survey, namely a POSS-I (Palomar
Observatory Sky Survey) plate from 1950, as well as in UKST (United Kingdom Schmidt Telescope)
infrared and red observations from 1981 and 1992. On all three plates we detected by eye inspection
a faint object, located approximately 18$\,$arcsec north of HD\,141272, which was not detected by
the SuperCOSMOS machine due to its small angular separation to the much brighter star and due to
its overlap with the diffraction spike (Fig. \ref{bild2}).

The diffraction spike of HD\,141272 intersects the northern object on all three plates hence, the
detection of this object would be inaccurate by means of most common detection techniques.
Nevertheless, we obtained a position measurement of the companion candidate on the POSS-I plate,
using the Source Extractor package (Bertin \& Arnouts 1996), included in the Starlink application
GAIA (Gray et al. 2004). The source extractor uses thresholding and deblending of point-spread
functions hence the method is more accurate than other detection techniques (e.g. Gaussian fitting)
under the circumstances in Fig. \ref{bild2}. However, an systematical error is possible, due to the
perturbation of the primary's spike. This error is larger in right accession than in declination
and would affect the measurement of the position angle rather than the separation (see section
\ref{data2}, Fig. \ref{rm3}), due to the orientation of the system (Fig. \ref{bild2} and
\ref{bild}).

Due to its brightness HD\,141272 saturates the POSS-I plate. Furthermore the PSF (point spread
function) is contaminated by the stray light of the companion candidate hence, position measurement
via PSF centering does not work sufficiently. We used the diffraction spikes of the saturated
primary to determine its position, since they are unaffected by the companion. We determined the
intensity center of a spike taking $\sim 30$ measurements for each spike using the data reduction
and analysis package ESO-MIDAS. The application of a linear regression gives the position of the
star as intersection of the two spikes and leads to very small astrometric uncertainties ($\Delta
\alpha_H = 0.047\,$arcsec and $\Delta \delta_H = 0.050\,$arcsec).

In addition to the detection on the POSS-I plate \newline HD\,141272 and its companion-candidate
are also detected in 2MASS images from observing epoch 2000. The 2MASS point source catalog (Cutri
et al. 2003) lists the position of both objects with accurate astrometric precision, see
Tbl.\,\ref{datatab}.

\begin{figure}
\resizebox{\hsize}{!}{\includegraphics{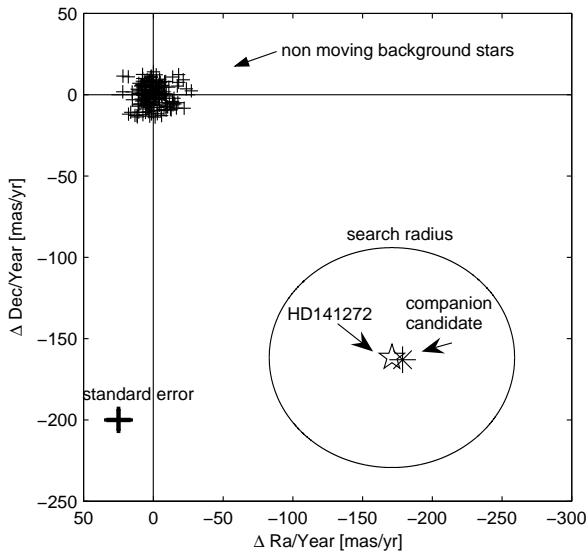}} \caption{Proper motion plot of HD\,141272
(cross) and its companion candidate (circle) and non moving background stars (upper left). X- and
Y-axis show the change of the positions (in mas/yr). The plot is based on POSS-I Schmidt plate (17
June 1950) and 2MASS catalog data (29 April 2000). Error estimates are taken as 2-$\sigma$ errors
from the background stars. Data points lying outside the background stars and outside a 5-$\sigma$
vicinity of HD\,141272 (large ellipse) are omitted, since these are either false detections or high
proper motions stars moving in other directions. The statistical error of all data points is shown
by the thick error cross in the lower left. The diagram shows the common proper motion of
HD\,141272 and its new companion with a confidence of $\sim 13 \sigma$.} \label{pm1}
\end{figure}

Equipped with these data  we determined the proper motion of all stars in a 15\,arcmin box around
HD\,141272 which are detected at the POSS-I plate and listed in the 2MASS point source catalog (see
Fig.\,\ref{pm1}). We derived the proper motion of all stars in the field by comparing the positions
of all detected objects. The majority of sources only shows small proper motion following a normal
distribution, since these stars are most probably at high distances. Using the Lilliefors test for
normal distribution we derived the subsample of stars belonging to the background stars, since
their proper motion follows a normal distribution (non moving background stars). The standard
deviation of the background stars gives the statistically derived proper motion error
($\sigma_{p.m.,\,\alpha}=8.8$\,mas/yr,\,\,\, $\sigma_{p.m.,\,\delta}=6.8$ mas/yr). Objects not belonging
to the background stars are considered as companion candidates, if they are lying within a
5-$\sigma$ vicinity of HD\,141272 (ellipse in Fig.\,\ref{pm1}). Other objects are omitted, since
these are either false detections or high-proper-motion stars moving in other directions.

The proper motion of the nearby star HD\,141272 is clearly separated from the background stars. The
companion candidate clearly shares the proper motion of HD\,141272 and will be denoted
HD\,141272\,B, hereafter. Fig.\,\ref{pm1} shows with high confidence ($\sim 13 \sigma$) that
HD\,141272\,A and B are co-moving over roughly 50 years. Due the above discussed astrometric
uncertainties of HD\,141272\,B this analysis gives a first indication of a new nearby young double
star system

Moreover, we used the non-moving background stars to estimate the positional error of the
detections in the POSS-I plate. The mean of the distribution shows the systematic error of the
POSS-I measurements ($\Delta_{\mathrm{sys},\,\alpha}=-4.5\,$mas/yr and
$\Delta_{\mathrm{sys},\,\delta}=-4.9\,$mas/yr as offset to $(0,0)$. The whole set of data points in
Fig. \ref{pm1} is shifted by that offset to correct for calibration errors between POSS-I and 2MASS
data. The standard deviation shows the statistical measurement error
($\Delta_{\mathrm{stat}}=\sigma_{p.m.}$) hence, can be applied as standard detection error. The
total detection error derived for the POSS-I plate is $\Delta \alpha=0.29\,$arcsec and $\Delta
\delta=0.25\,$arcsec. The additional systematic error for the companion candidate due to the
diffraction spike of HD\,141272 is not included in this error analysis.

\section{Follow-up observations}\label{data2}
\begin{figure}
\resizebox{\hsize}{!}{\includegraphics{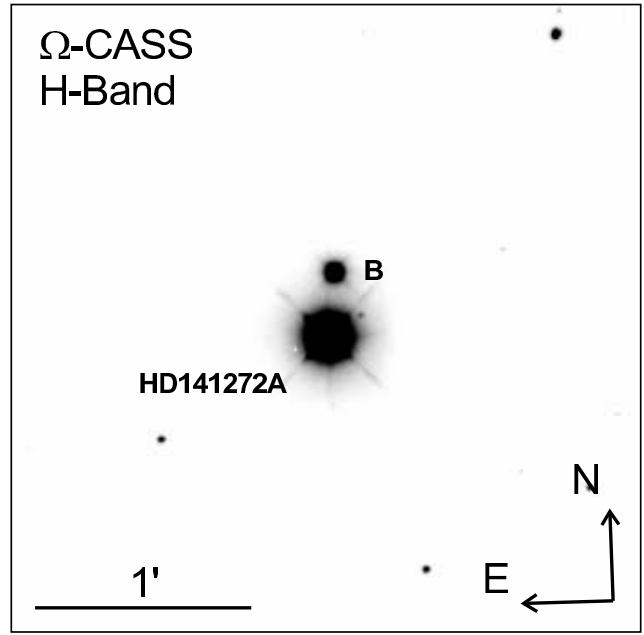}} \caption{H-band image of HD\,141272 and its
companion candidate taken with the near infrared camera $\Omega$-Cass at the 3.5\,m telescope of
the Calar Alto observatory in Spain.  The separation between HD\,141272 and its companion candidate
is $\sim 17.8$\,arcsec at a position angle of $\sim 352.62^{\circ}$ with a pixel scale of $\sim
0.2\,$arcsec/pixel. Note that HD\,141272 is slightly saturated.} \label{bild}
\end{figure}

In order to get a third epoch on our astrometric result and to detect or rule out further
companions we observed HD\,141272 again in April 2006 (Fig. \ref{bild}). We carried out H-band as
well as narrow-band observations ($1.644\,\mu$m) with the near infrared camera $\Omega$-Cass,
installed at the Cassegrain focus of the 3.5\,m telescope of the Calar Alto observatory in Spain.
$\Omega$-Cass is equipped with a 1024 $\times$ 1024 HgTeCd-detector with a pixel scale of $\sim$0.2
arcsec per pixel. We always used the shortest possible detector integration time (0.84\,s) to limit
strong saturation effects due to the bright star. For background subtraction we applied the
standard jitter technique and chose 12 jitter-positions. On each jitter position 49 integrations
(0.84\,s) were co-added, yielding a total integration time in the H-band of 8.2\,min. All images
were flatfielded with a skyflat image taken during twilight. The whole data reduction (background
subtraction, flatfielding, and shift+add) was carried out with the ESO data-reduction package
Eclipse (Devillard 2001).

\begin{table*}[t]
\caption{Separation and position angle of the co-moving companion HD\,141272\,B relative to its
primary HD\,141272\,A for all observing epochs. We also show the expected change of separation and
position angle in case that the companion is a non-moving background source, derived with the well
known proper and parallactic motion of the primary.} \centering \label{datatab}
\begin{tabular}{lccccccc}
\hline \hline
epoch & telescope/ & pixel scale & band & sep$_{obs.}$  & sep$_{if back}$ & PA$_{obs.}$ & PA$_{if back}$\\
$[$dd/mm/yyy$]$ & catalogs & $[$arcsec$]$ & & $[$arcsec$]$ & $[$arcsec$]$ & $[^{\circ}]$ & $[^{\circ}]$\\
\hline
17/07/1950 & POSS-I    & 1.0 & E (6442\AA) & 17.85$\pm$0.31   & $-$              & 353.6$\pm$1.1   & $-$\\
29/04/2000 & 2MASS     & 0.7 & JHK$_{S}$   & 17.83$\pm$0.150  & 26.92$\pm$0.33   & 352.42$\pm$0.48 & 14.61$\pm$0.75\\
20/04/2006 & 3.5m CA   & 0.2 & H           & 17.851$\pm$0.041 & 28.12$\pm$0.31   & 352.62$\pm$0.18 & 16.48$\pm$0.68\\
\hline
\end{tabular}
\end{table*}

We calibrated our $\Omega$-Cass image for relative astrometry, using the well known binary systems
HIP\,63322 and HIP\,82817, which we observed during the same night and with the same
instrumentation as our science image. Using the Hipparcos astrometry (Perryman et al. 1997) and
considering the maximal orbital motion of the calibration binaries we estimated the pixel scale
($192\pm 0.43\,$mas/pixel) and the orientation ($-1.86\pm0.18\,^{\circ}$) of the $\Omega$-Cass
images. This yields to the relative astrometric parameters of the system (Tbl. \ref{datatab}). For
the detection of both objects we used the Gaussian centroiding technique, implemented in ESO-MIDAS.

Further co-moving companions could be ruled out around HD\,141272 within an angular separation of
$\sim 5$ to $73\,$arcsec (1500\,AU of projected separation) with H-band magnitudes down to
18.3\,mag (S/N$=3$).

\begin{figure}
\resizebox{\hsize}{!}{\includegraphics{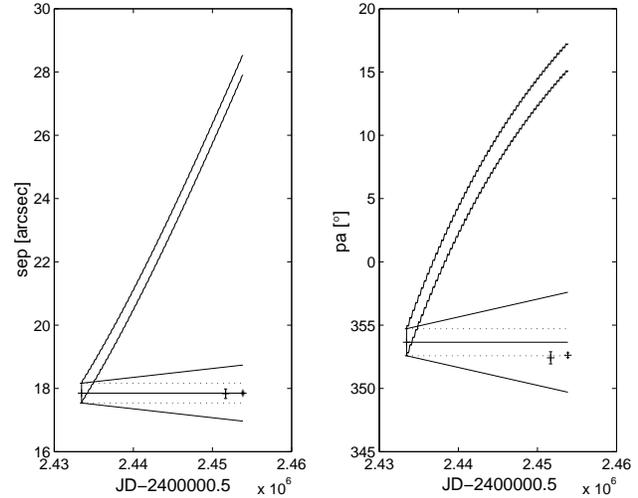}} \caption{Separation (sep) and position
angle (PA) for HD\,141272\,A and B from 1950 to 2006 (three data points). Upper lines show the
changes of the properties under the assumption HD\,141272\,B was a background star (including
parallactic motion of A) while the straight, opening lines give the range of the binary movement,
considering maximal orbital movement.  While the separation stays approximately constant there is a
change in the position angle, caused by the perturbation of the companions PSF due to the
diffraction spikes of the primary.} \label{rm3}
\end{figure}

HD\,141272\,A and B are separated by $\sim$\,17.8\,arcsec (Fig. \ref{bild}), hence the projected
separation of the system is approximately 380\,AU and its orbital period can be estimated with
Kepler's third law to be roughly 7000 years (we use 0.83\,M$_{\odot}$ for HD\,141272\,A and
0.26\,M$_{\odot}$ for B). During 56 years of epoch difference between the POSS-I and our H-band
observation, this yields maximal orbital motion as large as $\sim$0.5\,arcsec in separation
(edge-on orbit assumed) or $\sim$3$^{\circ}$ in position angle (face-on orbit assumed). Therefore,
we derived the separation and the position angle of the companion for all three observing epochs
which are summarized in Tab.\,\ref{datatab}. These results are also visualized in Fig.\,\ref{rm3}.
Note that absolute calibrated astrometric data, derived for the POSS-I image as described in
section \ref{data}, as well as catalog data from the 2MASS catalog is used in Fig.\,\ref{rm3},
while the third epoch data is based on relative astrometry, hence the uncertainties of that data
point are significantly smaller.

While the separation between HD\,141272\,A and B did not change during 56 years, we found a slight
decrease of its position angle. This effect is most likely due to the perturbation of the
companions PSF by the diffraction spike of the primary (see section \ref{data} and Fig.
\ref{bild2}). Nevertheless Fig. \ref{rm3} ensures the companionship of HD\,141272\,B, since all
data points are lying within the given error bars of the first epoch.

\section{Photometry and Spectroscopy}\label{photo}
The infrared colors of both components of the new binary system HD\,141272\,AB are listed in the
2MASS point source catalog, i.e. accurate J, H, and K$_{\rm S}$ band photometry is available for
the primary and its co-moving companion, which is summarized in Tab.\,\ref{2MASSphoto}.
Additionally the I-band magnitude of both components ($m_I=8.59\pm 0.02\,$mag for A and
$m_I=10.572\pm 0.02\,$mag for B) is measured in the second release of the DENIS database, while the
accuracy for HD\,141272\,A is limited due to saturation effects, hence the given error is probably
underestimated.

\begin{table}
\caption{2MASS Photometry of HD\,141272\,A and B} \label{2MASSphoto}\centering
\begin{tabular}{cccc}
\hline \hline
Comp. & J & H & K$_{\rm S}$\\
     & $[$mag$]$ & $[$mag$]$ & $[$mag$]$ \\
\hline
A & 5.991$\pm$0.021 & 5.610$\pm$0.027 & 5.501$\pm$0.018\\
B & 9.298$\pm$0.020 & 8.725$\pm$0.055 & 8.456$\pm$0.023\\
\hline
\end{tabular}
\end{table}
In order to obtain also unsaturated images of the primary we observed the binary system with
$\Omega$-Cass in the FeII (1.644$\,\mu$m) narrow-band filter. Thereby, we used again the 12 point
jitter pattern but co-added 15 integrations (4\,s) per jitter position, yielding a total
integration time of 12\,min. The bright primary as well as its fainter co-moving companion are both
well detected in this narrow-band image and their fluxes did not exceed the linearity level of the
$\Omega$-Cass detector. Hence, we could use this image to derive the magnitude difference between
the primary star and its companion and obtained $\Delta H_{FeII}=3.166\pm0.005$\,mag, fully
consistent with the magnitude difference derived from the 2MASS data in H-band ($\Delta
H=3.115\pm0.061$\,mag.)

Furthermore we acquired a low-resolution optical spectrum with EMMI at the NTT on La Silla to
determine the spectral type of HD\,141272\,B and prove its common distance with HD\,141272\,A. The
spectrum was taken in RILD and REMD mode covering a wavelength of 400-900\,nm with a resolution of
$R \approx 3000$ at 600\,nm. The data reduction followed the standard procedure for low-resolution
optical spectra: After bias subtraction, flat fielding and wavelength calibration with a HeAr arc
spectrum we corrected for the instrumental response and for telluric features using a spectrum of
HR5501 taken at the same airmass as HD\,141272\,B.

\begin{figure}
\resizebox{\hsize}{!}{\includegraphics{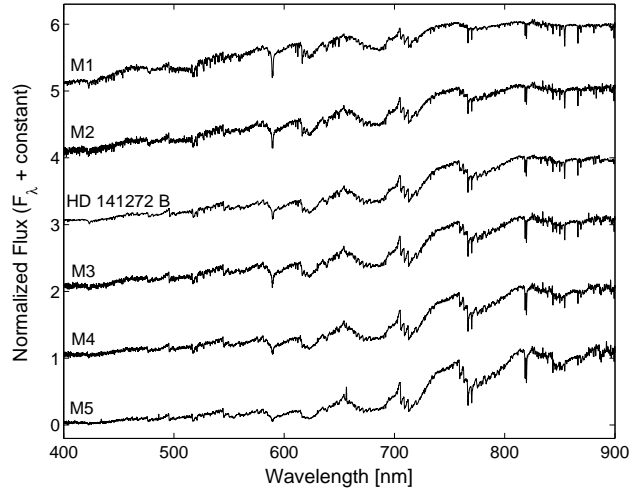}} \caption{Relative flux of the spectral sequence
from M1 to M5 (Bochanski et al. 2006) in comparison to the EMMI spectrum of HD\,141272\,B, ranging
from 400 to 900\,nm. The resolutions are comparable ($R\sim 3000$  for the EMMI spectrum and $R\sim
6000$ for the standard spectra at 600\,nm). HD\,141272\,B shows good agreement with an M3 star.}
\label{spektrum}
\end{figure}

We determined the spectral type by comparing our spectrum with a standard sequence of M dwarfs in
the same spectral range and with comparable spectral resolution (Boch\-anski et al. 2006), see Fig.
\ref{spektrum}. The best fit resulted in a spectral type of $M3.25 \pm 0.25$ which is consistent
with a spectral type of $M3.0 \pm 0.5$ determined from the TiO5 spectral index of 0.49 following
Cruz \& Reid (2002).

Adopting the latter spectral type as final we derived a spectrophotometric distance of $24.4 \pm
4.2\,$pc from the $M_J$ relation given in Cruz \& Reid (2002) and the $J$ magnitude from 2MASS,
assuming that the companion is on the Main sequence. The determined distance is in excellent
agreement with the HIPPARCOS measured distance of $21.35 \pm 0.48\,$pc for HD\,141272\,A,
confirming their common distance. Hence, we call the companion HD\,141272\,B.

\section{Conclusions}\label{concl}

With the astrometric data reduction and analysis techniques presented in this work, we could verify
the common proper motion of both components of the binary system HD\,141272\,AB during 56 years of
epoch difference between the first successful observation of this system on the POSS-I plates taken
in July 1950 and our H-band imaging obtained with $\Omega$-Cass in April 2006.

Furthermore we obtained an optical spectrum of the companion and derived its spectral type to range
between M2.5V and M3.5V. The infrared apparent magnitudes of the co-moving companion are fully
consistent with a M3 dwarf which is located at the distance of HD\,141272\,A which finally confirms
the companionship of this new binary system. The companion is an addition to the Catalog of Nearby
Stars within 25\,pc (Gliese \& Jahrei\ss\ 1991).

\begin{figure}
\resizebox{\hsize}{!}{\includegraphics{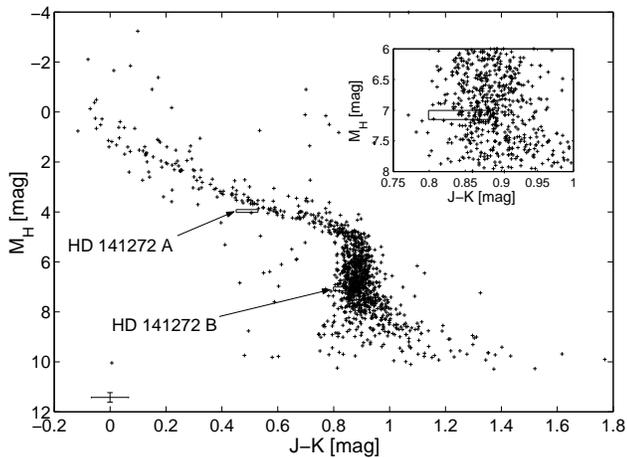}} \caption{J-K vs. $M_H$ diagram for the Pleiades and
HD\,141272\,A and B (rectangles symbolize the error boxes). The inserted plot shows HD\,141272\,B
and the surrounding Pleiades stars drawn to a larger scale. The main sequence of the cluster can be
seen although there are some outliers due to the mean distance module (5.97\,mag for Pleiades)
applied. The mean error of the Pleiades is shown by the error cross in the lower left.
HD\,141272\,A and B appear a little fainter than Pleiades stars of the same J-K color. This
indicates, that the system already reached the ZAMS.} \label{hrd}
\end{figure}

In order to get an estimation of the system age we compared the infrared photometry of
HD\,141272\,A and B with $\approx 1300$ members of the Pleiades cluster which are listed in the
WEBDA database (Mermilliod 1998). All objects are plotted in a J-K vs. M$_{\rm H}$ color-magnitude
diagram (Fig.\,\ref{hrd}). The colors of all objects are obtained from the 2MASS catalog and we
derived the absolute H-Band magnitudes of all comparison stars using their 2MASS H-band photometry
and a mean distance module of the Pleiades of 5.97$\,$mag (WEBDA database). The expected distance
uncertainty of the cluster members which results in an uncertainty of their absolute H-band
magnitudes was approximated with the angular diameter of the Pleiades cluster on the sky, assuming
a similar extension of the cluster also in the radial direction. The absolute H-band magnitudes of
HD\,141272\,A and B are derived with 2MASS photometry and the Hipparcos parallax of the binary
system. Compared to the Pleiades of the same J-K color HD\,141272\,A and B appear a little fainter,
indicating that the system is already on the ZAMS, which is similar to the results of earlier works
(Gaidos 1998; Wright et al. 2004).

If we assume that both components of the binary system have already reached the ZAMS we can
determine the mass of the secondary using equation (11) from Kirkpatrick \& McCarthy (1994) with
the given errors for the constants $a$ and $b$ and the range of the spectral type. We derived a
mass of
\[M_*=0.26^{+0.07}_{-0.06}\, M_{\odot}.\]

Future work should ascertain the age of the system and derive more properties of the M dwarf, which
enlarges the list of nearby low mass stars bound in binary systems.

\begin{acknowledgements}

We would like to thank the technical staff of the ESO NTT at La Silla as well as of the Centro
Astron\'omico Hispano Alem\'an (CAHA) at Calar Alto for all their help and assistance in carrying
out the observations.

In addition we would like to thank John Bochanski, Andrew West, Suzanne Hawley and Kevin Covey for
providing the electronic sequence of M-stars composite spectra.

T.O.B. Schmidt acknowledges support from a Thur\-ingian State Scholarship and from a Scholarship of
the Evangelisches Studienwerk e.V. Villigst.

This publication makes use of data products from the Two Micron All Sky Survey, which is a joint
project of the University of Massachusetts and the Infrared Processing and Analysis
Center/California Institute of Technology, funded by the National Aeronautics and Space
Administration and the National Science Foundation.

We use imaging data from the SuperCOSMOS Sky Survey, prepared and hosted by the Wide Field
Astronomy Unit, Institute for Astronomy, University of Edinburgh, \linebreak which is funded by the
UK Particle Physics and Astronomy Research Council.

This research has made use of the VizieR catalogue access tool and the Simbad database, both
operated at the Observatoire Strasbourg, as well as of the WEBDA database, operated at the
Institute for Astronomy of the University of Vienna.

The DENIS project has been partly funded by the SCIENCE and the HCM plans of the European
Commission under grants CT920791 and CT940627. It is supported by INSU, MEN and CNRS in France, by
the State of Baden-W\"urttemberg in Germany, by DGICYT in Spain, by CNR in Italy, by FFwFBWF in
Austria, by FAPESP in Brazil, by OTKA grants F-4239 and F-013990 in Hungary, and by the ESO C\&EE
grant A-04-046.

Jean Claude Renault from IAP was the Project manager.  Observations were carried out thanks to the
contribution of numerous students and young scientists from all involved institutes, under the
supervision of  P. Fouqu\'e, survey astronomer resident in Chile.

\end{acknowledgements}

\end{document}